\documentclass[epj]{svjour}
\usepackage{graphics,graphicx}
\usepackage{subfigure}

\newcommand {\be} {\begin{equation}} 
\newcommand {\ee} {\end{equation}} 
\newcommand {\Be}{\begin{eqnarray*}}
\newcommand {\Ee} {\end{eqnarray*}}
\newcommand {\bey} {\begin{eqnarray}} 
\newcommand {\eey} {\end{eqnarray}} 

\begin{document}

\title{The one-dimensional Lennard-Jones system:\\
collective fluctuations and breakdown of hydrodynamics \\}

\author{Stefano Lepri 
\thanks{\emph{Electronic address:} {\tt stefano.lepri@isc.cnr.it}}
\and Paolo Sandri  \and Antonio Politi 
}                     
\institute{ 
Istituto dei Sistemi Complessi, Consiglio Nazionale delle
Ricerche, Sez. Territoriale di Firenze \\
and Istituto Nazionale di Ottica Applicata, largo E. Fermi 6
I-50125 Firenze, Italy}

\date{Received: \today}

\abstract{ 
The dynamical correlations of a model consisting of particles constrained on the
line and interacting with a nearest--neighbour Lennard--Jones  potential are
computed by molecular--dynamics simulations. A drastic qualitative change of
the spectral shape, from a phonon--like to a diffusive form, is observed upon
reducing the particle density even ad moderate temperatures. The latter scenario
is due to the spontaneus fragmentation of the crystal--like  structure into an
ensemble of ``clusters" colliding among themselves. In both cases, the
spectral linewidths do not follow the usual $q^2$ behaviour for small
wavenumbers $q$, thus signalling a breakdown of linearized  hydrodynamics. This
anomaly is traced back by the presence of correlations due to the reduced
dimensionality.  
}

\titlerunning{The one--dimensional Lennard--Jones system}
\maketitle  

\section{Introduction}

Relaxation and transport phenomena in reduced spatial dimension ($D<3$) are
often qualitatively different from their three-dimensional counterparts. For
concreteness, imagine a large set of impenetrable spheres confined within a
narrow channel. If the mutual passage of particles is forbidden, the motion of
the spheres is necessarily correlated, even at long times, because the
displacement of a given particle over a long distance necessitates the motion
of many other particles in the same direction. This is a documented effect, for
example, in single-filing systems where particle diffusion does not follow
Fick's law~\cite{singlef,singlef2}. Another related instance is the enhancement
of  vibrational energy transmission in quasi-$1D$ systems like
polymers~\cite{morelli} or individual carbon nanotubes~\cite{nanotube}.

The distinguished signature of those effects is in the long-time behavior of
the associated correlation functions \cite{PR75}. As it is known, the latter
may display long--time tails leading to ill--defined transport coefficients or,
more generally, to the breakdown of customary hydrodynamics. Indeed, power--law
decay of correlations is expected to be a generic feature of one--dimensional
systems in presence of conservation laws ~\cite{NR02}. One important
consequence of such long--ranged correlations is that physical properties may
significantly depend on the system size and that the thermodynamic and
infinite--time limits may not commute. For instance, the tagged--particle
diffusion coefficient in $D=1$, that is finite for the infinite system is
found instead to vanish for a finite one (see \cite{pal} and references
therein). Another example is the divergence of the thermal conductivity
coefficient with the length observed in chains of anharmonic
oscillators~\cite{LLP97,LLP03} and hard--points gas~\cite{hpg}. The same type
of anomaly has been detected  also for a quasi--$1D$ model consisting of
spheres confined in a narrow channel~\cite{DN03}.

Computer simulations of simple toy models is an invaluable way of attacking
those problems. In particular, one would like to understand the conditions
under which those anomalies occur and possibly to classify the possible
universal features. In this paper we consider a simple model of point particles
constrained on the line and interacting with their nearest neighbours through a
Lennard-Jones force. This type of phenomenological interaction has been
throughly studied for decades by molecular--dynamics methods~\cite{fren}.
However its one--dimensional version has received little attention so far.
Previous studies focused on anharmonic effects~\cite{Cuc,schirm} and transport
properties~\cite{bazh}. In this respect, some evidence that the energy current
autocorrelation (the Green-Kubo integrand) shows a long--time tail of the above
mentioned type has been provided too \cite{MA88}. The same system has been also
proposed as a toy model to describe fracture nucleation~\cite{well,Oliv}. 

Of particular relevance in what follows is the work of Bishop and collaborators
\cite{Bishop1,Bishop2}. At variance with our model, they considered the  
case in which the interaction is extended to all particle pairs. They noticed
that the lifetime of long-wavelength fluctuations does not scale with their 
wavenumber $q$ as $q^{-2}$, as expected from standard hydrodynamics, but rather
as a nontrivial power $q^{-\mu}$. We anticipate that the results presented
henceforth are qualitatively consistent with theirs. However, the estimate of
the exponent $\mu=1/3$ given in Ref.~\cite{Bishop2} is significantly different
from the value found here, $\mu\simeq 1.5$, which is consistent with previous 
measurements in chains of coupled oscillators~\cite{L98,chaos}. This issue 
is relevant especially for assessments about the universality of the 
scaling laws. 

The present paper is organized as follows. In section 2 we define the model
and its physical parameters. Section 3 is devoted to a discussion of the
nature of the ground state at zero temparature. This is important to understand
the effect of density changes on the system dynamics. In Section 4 we present 
our simulation results and show how the low-density properties are 
related to the kinetics of particle clusters. Finally, we summarize our 
results in the concluding Section.  

\section{The model}

We consider an array of $N$ point-like identical atoms ordered along a line. 
The position of the $n$-th atom is denoted with $x_n$.  By fixing the mass,
without loss of generality, equal to unity and assuming that interactions are
restricted to nearest-neighbour pairs, the equations of motion write
\begin{equation} 
{\ddot x}_n = - F_n + F_{n-1} \quad ; \qquad 
F_n=- V'(x_{n+1} - x_n)  \, ,
\label{eqmot} 
\end{equation}  
where $V'(z)$ is a shorthand notation for the first derivative of the
the interparticle potential $V$ with respect to $z$. The particles 
are confined in a simulation ``box"
of length $L$ with periodic boundary conditions 
\begin{equation}
x_{n+N} \;=\; x_{n} \,+\, L  \quad .
\label{pbc}
\end{equation}
Accordingly, the particle density $d=N/L$ is a state variable to be considered
together with the specific energy (energy per particle) that will be denoted by
$e$.

In the present work we focus on the Lennard--Jones potential
that in our units reads
\begin{equation}
V(z) \;=\;  {1\over 12}
\bigg({1 \over z^{12}}\, -\, {2\over  z^6}\,+\, 1 \bigg)\quad .
\label{lj}
\end{equation}
For computational purposes, the coupling parameters have been fixed in such a
way as to yield the simplest form for the force. With this choice , $V$ has a
minimum in $z=1$
and the resulting dissociation energy is $V_0=1/12$. Notice that for
convenience we set the zero of the potential energy in $z=1$.  The presence of
the repulsive term in one dimension ensures that the  ordering is preserved
(the particles do not cross each other).  

Before closing this section, let us recall that most previous studies 
\cite{MA88,bazh,Bishop1,Bishop2} dealt with the case in which the interaction
is not restricted to nearest neighbours but rather extends all particle pairs.
In practice, as pointed out in Refs.~\cite{Bishop1,Bishop2}, even in this case
the interaction is limited to about 2-3 neighbours.  Therefore, we do not
expect that the results reported below will be significantly altered when
taking into account the interaction among all pairs. 

\section{The ground state at $T=0$}

In order to understand the physical features of the model let us briefly
discuss its equilibrium properties. Since the system is one dimensional  with
only nearest--neighbour and short-ranged interaction, no actual phase
transition at finite temperature, $T > 0$, can occur. Nevertheless, we will
see that the dynamics  of collective modes may strongly depend on the state
variables. Therefore, even if it is not legitimate to speak of ``phases"
in a strict  thermodynamic sense, it is sensible to distinguish between
regions where the dynamics is qualitatively different. 

\begin{figure}[ht!]
\includegraphics[clip,width=8cm]{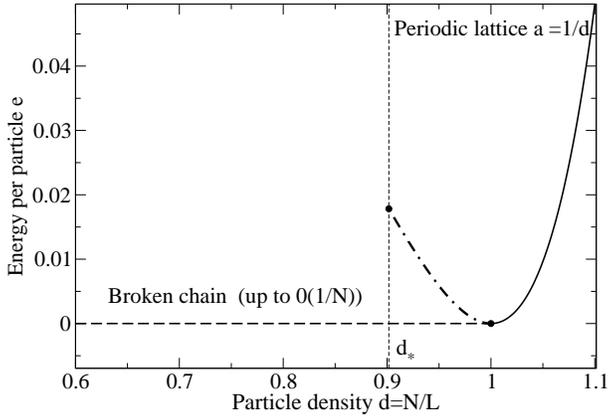}
\caption{The $T=0$ minimal--energy configurations of the Lennard--Jones
chain as a function of the particle density.} 
\label{fig:gstate}
\end{figure}

Let us start discussing the equilibrium configurations at $T=0$ as a function
of the particle density $d$. This issue, which is important for the choice of
the initial conditions in microcanonical simulations (see below), has 
been discussed in Ref.~\cite{Still} for the Lennard--Jones chain where the pair
interaction is among all particles. In view of the short--range nature of the
potential we do not find significative difference with the case at hand here.
For convenience, we summarize the  main results in Fig.\ref{fig:gstate}. Three
density regimes are distinguished: 
\begin{itemize}
\item For $d\ge 1$ the ground state consists of equally spaced particles 
at a relative distance $a=1/d$ (homogeneous solution). The chain is under 
compression and the total energy decreases upon decreasing $d$ up to the 
minimum value which is attained for $d=1$.
\item For $d_*<d<1$ the homogeneous solution (chain under tension) 
becomes a relative minimum. The ground state consists of equally spaced 
particles ad a distance approximatively $a\approx 1$  except for a couple 
which lies at a large distance. In other words, the minimal energy configuration
is attained by breaking the chain at one bond. At the critical value    
$d \;=\;d_* \;=\; \sqrt[6]{\frac{7}{13}}\;=\; 0.901971 \ldots$,
the homogeneous configuration undergoes an instability and disappears.
\item For $d<d_*$ the broken chain state is the unique minimal
energy configuration.
\end{itemize}
Notice that for $d<1$ the ground--state energy is basically 0 up to 
terms ${\mathcal O}(1/N)$ since only a couple of particles out of $N$ 
sits at a distance which is significantly larger than 1. Furthermore, as 
noted in Ref.~\cite{Still}, further fragmentation in two or more 
shorter chains cannot produce further energy minima.

The phase diagram of  Fig.~\ref{fig:gstate} suggests a first--order  phase
transition at $d=1$ where the second derivative of the energy  density with
respect to $d$ undergoes a jump discontinuity. As this quantity should be
proportional to the Young's modulus this physically means that the chain
undergoes a fracture and looses its elastic tension. 

At finite temperature we expect that these features should be washed out
but, as observed for other models~\cite{flach}, the remnants of the transition 
should somehow manifest themselves in the dynamics. 

\section{Dynamical correlations}

We have performed equilibrium microcanonical simulations by integrating
Eqs.~(\ref{eqmot}) (with periodic boundary conditions (\ref{pbc}) )  by means
of a fourth--order symplectic algorithm \cite{MA92}. For the different values
of density considered henceforth, initial  conditions were chosen to be in the
ground state described in the  previous section. The inital velocities were
drawn at random  from a Gaussian distribution and rescaled by suitable factors
to assign the kinetic energy to the desired value and to set the total initial
momentum equal to zero. A suitable transient is elapsed before acquisition of 
statistical averages. Conservation of energy and momentum was monitored during
each run. This check is particularly crucial at high energies/densities where
the strongly repulsive  part of the force comes into play and may lead to
significant inaccuracies. The chosen time--step  (0.005--0.05) ensures
energy conservation up to a few parts per  million in the worst case.   

We computed the dynamical structure factor, namely the square modulus of 
temporal Fourier transform of the particle density 
\begin{equation}
\rho(q,t)  \;=\; {1\over N}\,\sum_n \, \exp(-iqx_n) \quad ,
\label{dens}
\end{equation}
which is defined as
\begin{equation}
S(q,\omega) \;=\; 
\big\langle \big| \rho(q,\omega )\big|^2 \big\rangle  \quad .
\label{strutf}
\end{equation}
The square brackets denote an average over a set of independent
molecular--dynamics runs. By virtue of the periodic boundaries, the allowed
values of the  wavenumber $q$ are integer multiples of $2\pi/L$.
The reliability of the spectra have been checked against different
choices of the run durations and sampling times. In some particular
cases we also verified that the results are not affected by 
data windowing that, in principle, may affect the measured linewidths. 

\subsection{The $d=1$ case}

In this case we expect a ``lattice--like" behaviour with each particle
oscillating around its equilibrium positions. In other words, we can 
introduce the change of coordinates $x_n=u_n+na$ where $a=1/d$ is the 
constant lattice spacing. The computation of $S(q,\omega)$ 
can be more conveniently performed by resorting to the 
collective coordinates
\begin{equation}
U(q) \;=\;{1\over \sqrt{N}} \sum_{n=1}^N \, u_n \, \exp({-iqn})
\quad q={2\pi k \over Na} 
\label{Uk}
\end{equation}
for $k$ being an integer comprised between $-N/2+1$ and $N/2$. This latter
expression is computationally more efficient than the definition
(\ref{dens}). In fact, standard Fast Fourier Transform routine can be used
to evaluate $U(q)$ (provided that $N$ is a power of 2). On the other hand, 
by expanding to the leading order in $q$, it is seen that 
the structure function (\ref{strutf}) is proportional to 
$q^2 \langle|U(q)|^2\rangle$. We checked numerically that this approximation
is very accurate in the range of $q$ values considered henceforth.

\begin{figure}[ht!]
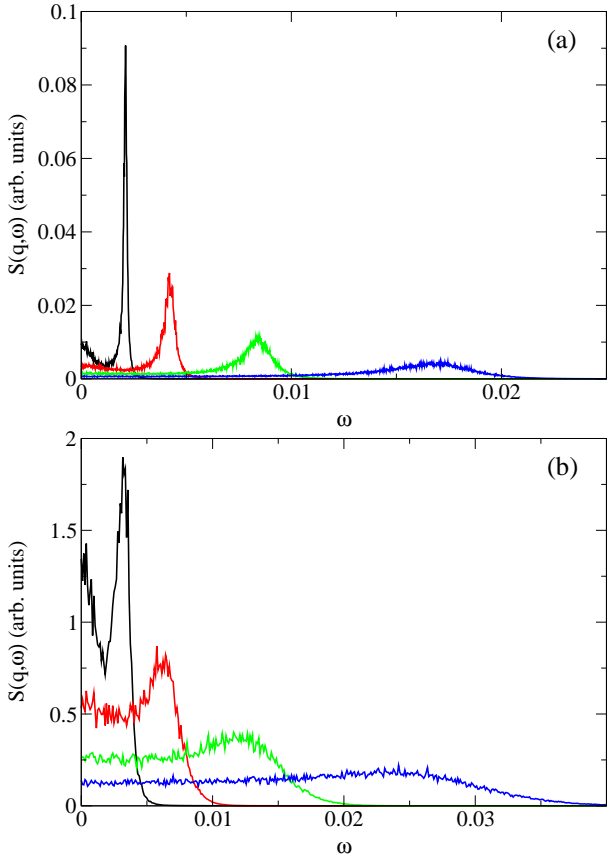

\includegraphics[clip,width=8cm]{Flut_N8192_d1.0_e0.02.eps}
\hspace{2mm}
\includegraphics[clip,width=8cm]{Flut_N8192_d1.0_e0.2.eps}
\caption{
Structure factors for $d=1$, $N=8192$ and different $q$s 
corresponding to indices $k=1, 2, 4, 8$ in eq.~(\ref{Uk}) (left to right). 
Microcanonical simulations are performed for the energy densities 
$e=0.02$ (a) and $e=0.2$ (b). The spectra are averaged over an ensemble 
of about 100  initial conditions.  } 
\label{fig:spectra1}
\end{figure}

In Fig.~\ref{fig:spectra1} we report two representative data sets for low
($e=0.02$) and high energies ($e=0.2$) compared with the well depth. The most
distinguished feature of the the spectra is a narrow phonon peak.  For the low
energy case, its frequency $\Omega$ agrees with the one computed by the
harmonic approximation of the Lennard-Jones potential:
\begin{equation}
\Omega(q) \;=\; 2\sqrt{V^{\prime\prime}(1)}\, \big| \sin\frac{qa}{2} \big|
\quad.
\end{equation} 
Indeed, the estimated sound speed $c=2.77\pm0.06$ is only slightly larger  the
value obtained from this latter expression ($c=\sqrt{6}\approx 2.49...$).  This
indicated that nonlinear terms are weak in this energy range. At higher
energies, the particles are less sensitive to the bottom part of the potential
well and the phonon peaks shifts to higher frequency.  The data in
Fig.~\ref{fig:spectra1}b correspond to $e=0.2$, more than 20 times the well
depth; comparing the values of Fig.~\ref{fig:spectra1}a it can be ascertained
that the frequencies are roughly 50 \% larger than in the previous case.

For small enough wavenumbers, besides the phonon peak, a small zero--frequency 
component appears whose relative weight increases at larger energies (see again
Fig.~\ref{fig:spectra1}). This feature, that has not been reported in previous
works on the Lennard-Jones chain \cite{Cuc}, is reminiscent of the well--known
triplet structure observed in fluids where a central Rayleigh peak is 
accompanied by two narrow Brillouin lines placed at $\pm cq$ \cite{forster}.
Moreover, according to the standard hydrodynamic theory, the ratio of the areas
under half the Rayleigh peak and one Brillouin peak is $C_p/C_V - 1$.  On the
other hand, the specific heats ratio is expected to be very close to unity for
a ``crystal-like" structure as the one we are facing in our simulation. This is
thus consistent with the intuitive idea that a sizeable  central component
should occur only when the systems is ``fluid enough" i.e. when fluctuations of
the particles' positions are large. This  is expected for large temperatures
and/or small densities \footnote{The existence of a central component also
means that there is a coupling between density and energy fluctuations which
are the hydrodynamic fields of our model. To support this interpretation we
computed also the spectra of the local energy field in some cases. They
do display a similar structure with a large component at the phonon
frequency. }. In the next section we will see that this expectation is further 
confirmed by the simulations.  

\begin{figure}[ht!]
\includegraphics[clip,width=8cm]{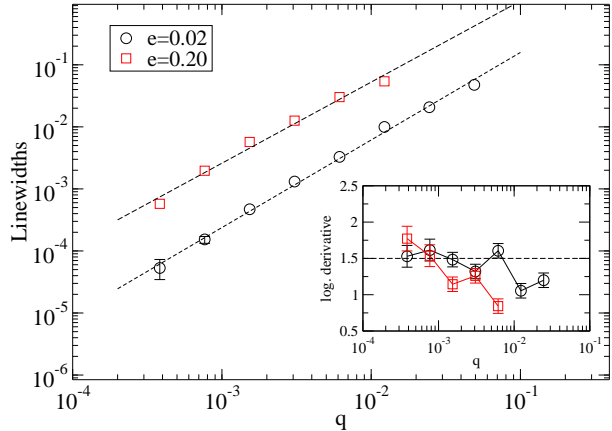}
\caption{The linewidths of the phonon peak for $d=1$, $e=0.02$
and $e=0.2$. Error bars are reported only when significantly larger than
the symbols. The inset shows the logarithimc derivative evaluated 
by finite differences.} 
\label{fig:gamma1}
\end{figure}

In Fig.~\ref{fig:gamma1} we report the linewidths of the Brillouin peaks as a
function of the wavenumber $q$ and for the two energy values chosen above.
The data reported there correspond to those wavenumbers for which the 
linewithds are small enough with respect to the peak frequency (typically less
the one tenth) to allow for a meanigful estimate. The linewidths are computed
by evaluating the frequencies at which the spectrum is  one half of its maximal
value. In the cases in which the spectral resolution is not high enough we 
used by a Lorenzian fit to improve the accuracy. Incidentally, we noticed
that the the fitting is reasonably good only in the peak region while 
substantial deviations on the tails are observed.

Remarkably, the linewidths do not scale as $q^2$ as expected from standard
hydrodynamics but rather as $q^\mu$. For $e=0.02$ our best--fit estimate is
$\mu= 1.47\pm0.05$. To better appreciate the reliability of this value, in the
inset of Fig.~\ref{fig:gamma1} we report the logarithmic derivative of the 
data that show a plateau around 1.5. The data for $e=0.2$ are instead 
less conclusive. Although a power law fit still yields a comparable 
value ($\mu= 1.41\pm0.07$) the plot displays a residual curvature that
indicate some relevant subleading corrections. 

\subsection{The $d<1$ case}

\begin{figure}[ht!]
\includegraphics[clip,width=8cm]{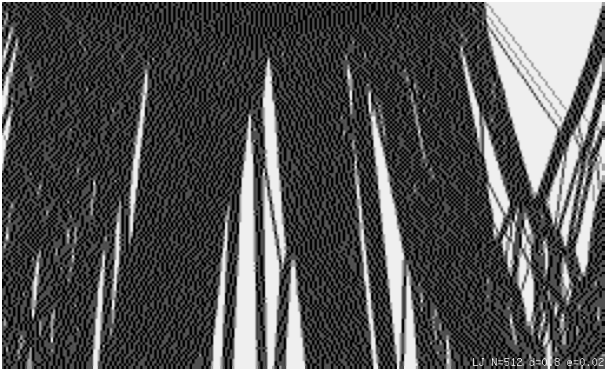}
\caption{The evolution of the particle density for $d=0.8$, $N=512$ and 
$e=0.02$. Along the horizontal direction a black pixel is drawn at each 
particle location. Time increases downwards.} 
\label{fig:clust}
\end{figure}

At finite temperature, when the density is lowered below $d=1$, the periodic 
ground state destabilizes and this leads to a completely different dynamics. To
avoid the additional features connected with the presence of the metastable
branch (see Fig.~\ref{fig:gstate}) we consider the case $d<d_*$.
Fig.~\ref{fig:clust} shows the evolution of the particle density field starting
from the minimal-energy configuration for $d=0.8$. After a transient, the system
spontaneously fragments in a series of clusters of different lengths.  Within
each cluster, the particles are separated by an average distance equal to the
equilibrium distance of the Lennard-Jones potential (1 in our units).

The relaxation is thus ruled by the clusters' kinetics. The latter
consists of two different processes, fragmentation and collisions whose
characteristic times we denote by $\tau_F$ and $\tau_C$ respectively. Let
us start giving an estimate of the average number $m$ of particles in each
cluster. For convenience, we write $m=1/f$ where $f$ is the average fraction of
broken bonds. The average time between collisions is given by the ratio between
the typical clusters' separation and the typical speed $v$. A straightforward
calculation yields 
\begin{equation}
\tau_C \;=\; \frac{1}{vf}\, \Big(\frac{1}{d} - 1\Big) 
\end{equation} 
(remember that the equilibrium distance is 1 in our units). On the other hand,
we expect fragmentation to be a thermally activated process with an activation 
energy equal to the well depth $V_0$. Accordingly, 
\begin{equation}
\tau_F \;=\; \tau f \, \exp{\beta V_0}  \quad, 
\label{arre}
\end{equation}
where $\tau$ is a suitable prefactor setting the typical ``attempt time" for 
the process and $\beta$ is the inverse temperature. The condition for 
kinetic equilibrium is obtained by letting $\tau_C = \tau_F$,
\begin{equation}
f \; =\; \sqrt{1-d \over \tau vd }\,{\exp\Big( -{\beta V_0 \over 2}\Big)}\quad.
\label{bbonds}  
\end{equation}
Notice that, as a consequence of this estimate, the characteristic
time is proportional to $\sqrt{1/d - 1 }\,\exp(\beta V_0/2)$
and that thermalization may become very slow for low temperatures and 
densities.

\begin{figure}[!ht]
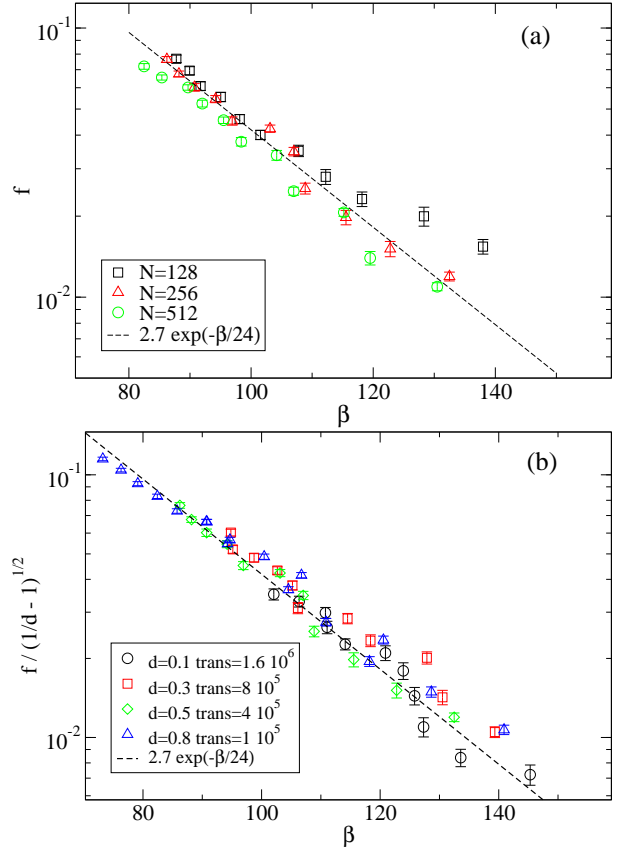

\includegraphics*[clip,width=8cm]{Dens_Brock_N_d0.5.eps}
\hspace{2mm}
\includegraphics*[clip,width=8cm]{Dens_Brock_T.eps}
\caption{The average fraction of broken bonds $f$ as a function 
of the inverse kinetic temperature $\beta$; (a) dependence of $f$ on the 
number of particles $N$ for fixed density $d=0.5$; (b) check 
of the formula (\ref{bbonds}) for fixed $N=256$.} 
\label{fig:dbrock}
\end{figure}

This prediction is in fairly good agreement with the numerical data (Fig.
\ref{fig:dbrock}). There, we plot the average fraction of broken bonds $f$
as a function of the inverse kinetic temperature.
The quantity $f$ is measured by counting at each time  the number of pairs whose
distance is larger than some prescribed threshold that we fixed equal to 1.5
(some 50\% above the inflection point of the Lennard--Jones potential). From
the  above discussion, it is clear that  the choice of well--thermalized
initial conditions is crucial. Indeed we found that the time to reach the
equilibrium value of $f$ increases (about linearly) with the number of
particles.   In Fig.~\ref{fig:dbrock}a it is shown that $f$ is an extensive
parameter  whose equilibrium value can be evaluated already for systems of a 
few hundred particles. In Fig.~\ref{fig:dbrock}b we check that the  scaling
behaviour predicted by the kinetic argument reported above  is in agreement
with the simulation data. Notice in particular that the  factor 1/2 in the
exponential term is very well accounted for by the  data and that the
prefactors display a rather weak dependence on $\beta$.      

\begin{figure}[!ht]
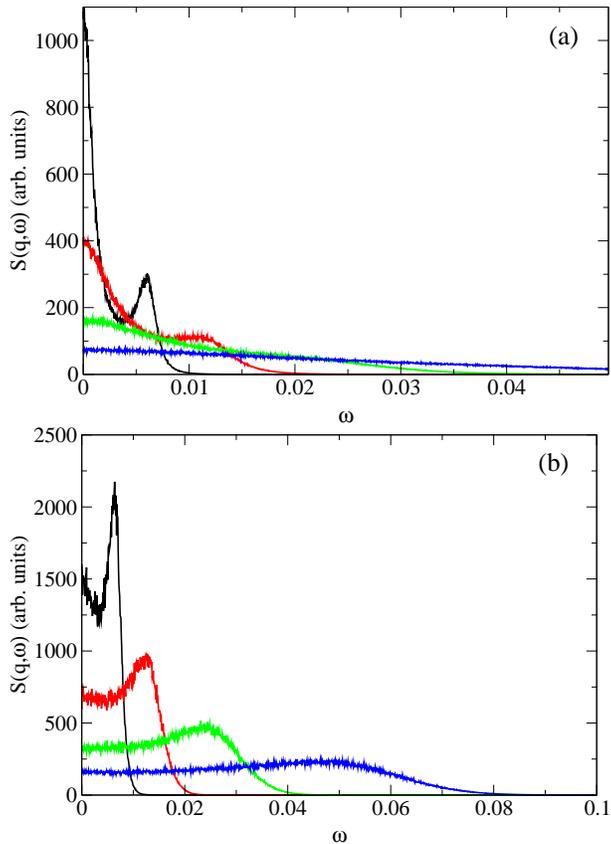

\includegraphics*[clip,width=8cm]{Flut_N1024_d0.8_e0.02.eps}
\hspace{2mm}
\includegraphics*[clip,width=8cm]{Flut_N4096_d0.8_e0.2.eps}
\caption{
Structure factors for $d=0.8$ and different $q$s 
corresponding to indices $k=1, 2, 4, 8$ in eq.~(\ref{Uk}) (left to right). 
Microcanonical simulations are performed for the energy densities 
$e=0.02$, $N=1024$ (a) and $e=0.2$, $N=4096$ (b). The spectra are averaged over an ensemble 
of about 500 and 100  initial conditions respectively.
} 
\label{fig:spectra2}
\end{figure}

In view of the above dynamical features, we expect that the  the spectra of
density fluctuations should be qualitatively different from the $d=1$ case.
This is confirmed by the simulation reported in  Figs.~\ref{fig:spectra2} (in
this low-density regime we use the  definition (\ref{strutf}) of
$S(q,\omega)$). Indeed, even at low energies, a large central component appears
which we associate with the diffusive behaviour of the clusters. The secondary
Brillouin peaks occur only for small enough wavenumbers and are
weaker:  for the smallest wavenumber shown in Fig.~\ref{fig:spectra2} the
secondary peak has about 40\% spectral power of the central component. This
observation is consistent  with the intuitive expectation that at lower
densities the system should respond like a fluid.

The behavior for $e=0.2$ is shown in Fig.~\ref{fig:spectra2}b. The structure
factor is qualitatively similar to the case of  Fig.~\ref{fig:spectra1}b. This
is expected, since in the high energy  limit the model approach the case of
hard points colliding elastically.  It is thus plausible that the dynamics is
practically independendent on the particle density. Notice also that in this
limit the model  becomes an almost--integrable dynamical system and some  some
pathological behaviour (e.g. slow thermalization) may be expected.

\begin{figure}[ht!]
\includegraphics[clip,width=8cm]{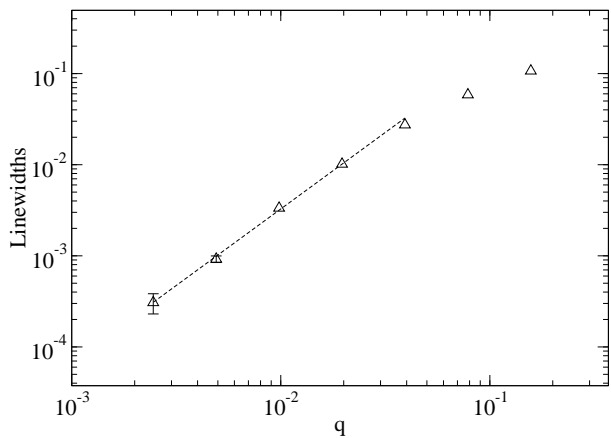}
\caption{The linewidths of the central peak for
$d=0.8$ and $e=0.02$. Error bars are reported only when significantly 
larger than the symbols. } 
\label{fig:gamma2}
\end{figure}

In analogy with the case $d=1$ we analysed the dependence of the spectral
linewidths on the wavenumber, considering only those $q$ values that correspond
to narrow enough lines. Fig.\ref{fig:gamma2} shows the half-widths of the
central (Rayleigh) peak as a function of the  wavenumber $q$ for the $e=0.02$.
In the accessible range we find  again a nontrivial scaling
$q^\mu$ with $\mu = 1.60\pm 0.05$.

\section{Conclusions}

Our numerical analysis of the one-dimensional Lennard-Jones system has shown
that long--ranged correlations induced by the reduced spatial dimensionality
may significantly affect the density fluctuations. 
The most striking feature is the anomalous scaling of the Rayleigh and
Brillouin peak widths in the hydrodynamic limit, $q \to 0$. The physical
meaning of this behaviour is  understood by recalling that, in the standard
framework, the linewidths  are connected to transport
coefficients~\cite{forster}. For instance the width of  the Brillouin peak is
$\Gamma q^2$ where $\Gamma$ is the sound attenuation constant. The anomalous
scaling can be recasted in terms of a  wavenumber--dependent constant
$\Gamma(q) \sim q^{\mu-2}$. Since our simulations clearly indicate that
$\mu<2$, this implies that $\Gamma$  {\em diverges} in the $q\to 0$ limit. In
other words, a long-wavelength  disturbance is damped on a typical distance
which becomes very large. In some sense, one may think of this as a
superdiffusive process, intermediate between standard diffusive and  ballistic
propagation. Our result is thus closely related to  the analysis performed in
Ref.~\cite{denis} for the hard--point gas. 

This feature has been previously observed also in other one--dimensional
lattice models~\cite{L98,chaos}.  A similar anomalous behaviour of the
viscosity of a $1D$ lattice gas has also been reported in Ref.~\cite{DLQ89}.
However, the exponent $\mu$ found in the present work is about 10\% smaller
then those measured in Refs.~\cite{L98,chaos} (1.5 against 1.67).  Remarkably,
both values  are very close to theoretical estimates presented previously in
the literature. Ernst \cite{E91} claimed that $\mu=5/3$, while the more refined
mode-coupling analysis by Wang and Li \cite{WL04} gave $\mu=3/2$ for the
specific model they considered. We anticipate that this latter value is
actually supported by numerical solution of the mode-coupling
equations~\cite{delfini} that, according to Ref.~\cite{schirm}, should describe
the dynamical properties of our model. This estimate is in excellent agreement
with the data reported above. Whether the small differences in the exponents are
due to numerical errors and subleading corrections or they indicate the
existence of two  different ``universality classes" is still an open question.

For $d=1$ the system retains the dynamical features of a one-dimensional
crystal. Of course, by virtue of the short-ranged interactions, the model
cannot have a genuine solid phase (Landau-Peierls instability). Nonetheless,
we have found that for a finite system we can still reason in terms of an
effective phonon dispersion and damping. In the low-density region ($d<1$) we
have shown how the relevant time scales are instead dictated by the collision
and recombination processes of one-dimensional ``clusters" of particles. Such
processes may lead to a remarkable increase of the thermalization times
when the kinetic temperature becomes smaller than the binding energy
(see Eqs.~(\ref{arre} and \ref{bbonds})).

In the present study we limited ourselves to the case in which the  system is
initialized close to the ground state. It would be interesting to examine the
relaxation from the metastable state which exists for  $d_*<d<1$. In this 
respect there is a connection with the work of Oliveira \cite{Oliv} that
studied the fracture nucleation in the same model. At variance with  what
discussed here, he starts from the uniformly stretched chain, namely from the
metastable state illustrated in the phase diagram of Fig.~\ref{fig:gstate}. He
founds that the breaking time is orders of magnitude longer than what expected
from Kramers-type estimate. This is  presumably to be traced back to subtle
long-ranged correlations of the very same type studied here.

To conclude, we wish to mention the fact that the anomalous scaling 
may manifest also at the level of energy transport in this model.
Indeed, nonequilibrium simulations show that the thermal conductivity
coefficient diverges with the system length also in the 
low--density regime \cite{cip}. This is a further evidence that 
nonequilibrium processes in one dimension are peculiar and deserve 
a special attention.

\section*{Acknowledgments}

This work is supported by the PRIN2003 project
{\it Order and chaos in nonlinear extended systems} funded by MIUR-Italy.
Part of the numerical simulation were performed at CINECA supercomputing
facility through the INFM {\it Iniziativa trasversale ``Calcolo Parallelo"}
entitled  {\it Simulating energy transport in low-dimensional systems}.

\end{document}